# Balancing Security and Privacy: The Pivotal Role of AI in Modern Healthcare Systems

Deepthy K Bhaskar[1], Dr. Minimol B[2] and Dr. Binu V P[3]

[1] Model Engineering College, APJ Abdul Kalam Technological University,Thiruvananthapuram, Kerala, 695016, India
[2] Model Engineering College, APJ Abdul Kalam Technological University,Thiruvananthapuram, Kerala, 695016, India
[3] Model Engineering College, APJ Abdul Kalam Technological University,Thiruvananthapuram, Kerala, 695016, India

`dmdl22jan005@mec.ac.in`

**Abstract.** As digital threats continue to grow, organizations must find ways to enhance security while protecting user privacy. This paper explores how artificial intelligence (AI) plays a crucial role in achieving this balance. AI technologies can improve security by detecting threats, monitoring systems, and automating responses. However, using AI also raises privacy concerns that need careful consideration.We examine real-world examples from the healthcare sector to illustrate how organizations can implement AI solutions that strengthen security without compromising patient privacy. Additionally, we discuss the importance of creating transparent AI systems and adhering to privacy regulations.Ultimately, this paper provides insights and recommendations for integrating AI into healthcare security practices, helping organizations navigate the challenges of modern management while keeping patient data safe.

## 1   Introduction

In our digital world, organizations are collecting and using more data than ever. While this offers many benefits, it also brings challenges, especially regarding security and privacy [1]. As companies rely on technology to manage sensitive information, they must ensure strong security measures are in place to protect against threats. At the same time, people are more aware of privacy issues, making it essential for organizations to keep personal data safe. Artificial intelligence (AI) has become a valuable tool in addressing these challenges. AI can help improve security by detecting potential threats, monitoring systems, and responding quickly to breaches. However, using AI also raises concerns about privacy, including how data is used and the transparency of AI decisions.

The healthcare industry has been at the forefront of adopting new technologies, with Artificial Intelligence (AI) emerging as a powerful tool for transformation. AI-driven systems have the potential to revolutionize several areas of healthcare, from diagnosis to treatment decisions and patient monitoring [2]. Integrating AI into clinical practice can enhance the accuracy, efficiency, and accessibility of healthcare services, ultimately improving patient outcomes [3]. However, the use of AI in healthcare also brings significant concerns about data security and patient privacy. Healthcare data, particularly electronic health records, are extremely sensitive and valuable, making them prime targets for cyber-attacks and data breaches [4]. To ensure AI is used responsibly and ethically in healthcare, it is crucial to address these concerns and balance the advantages of AI with the need to protect patient privacy and data security.

## 2    Background

The healthcare industry generates vast amounts of data, much of which is sensitive in nature, including patient medical histories, diagnoses, treatments, and personal identification information. As healthcare systems have transitioned from paper-based to electronic health records (EHRs), managing, securing, and leveraging this data has become a top priority. This shift has opened the door for significant technological advancements, particularly in the fields of artificial intelligence (AI) and machine learning, which offer the potential to analyze healthcare data at an unprecedented scale and depth. AI has shown promise in transforming numerous aspects of healthcare. From assisting in diagnostics to supporting treatment planning and improving patient monitoring, AI-driven solutions are being integrated into clinical workflows to improve care quality and efficiency. For example, AI algorithms have been developed to analyze medical imaging, predict disease progression, and personalize treatment plans based on a patient's unique data. These innovations have the potential to enhance clinical outcomes, reduce costs, and increase access to healthcare services.

This paper explores how AI can be used to enhance both security and privacy in healthcare, focusing on the pivotal role of technologies such as encryption and privacy-preserving machine learning techniques. It also considers how organizations can adopt AI to improve healthcare delivery without compromising the confidentiality and integrity of patient data.

### 2.1    The Role of AI in Enhancing Healthcare Security

AI plays a vital role in strengthening security in the healthcare sector. It can address growing threats to data protection in several ways:

- **Threat Detection and Response:** AI monitors healthcare networks in real time, identifying unusual activities and potential cyber-attacks. Machine learning helps detect vulnerabilities and respond faster than traditional systems [5].
- **Automated Security:** AI automates tasks like monitoring network traffic and flagging suspicious access attempts [6] , reducing the load on human teams. This continuous surveillance allows for quick, automated responses to potential threats..

- **Fraud Detection:** AI can spot patterns in large datasets, helping to detect fraudulent activities in healthcare billing and insurance [7] , such as false claims or unauthorized access.
- **Encryption and Secure Data Sharing:** AI supports encryption of patient data, ensuring its protection during storage and transmission [8]. It also enables secure sharing of data between healthcare organizations, ensuring privacy is maintained during exchanges..

## 2.2 Privacy Preserving Techniques in AI for Healthcare

This section would cover the methods and technologies that allow healthcare organizations to use AI while safeguarding patient privacy. It could include:

- **Differential Privacy:** How differential privacy adds noise to datasets, allowing AI models to learn useful patterns without exposing individual patient information [9].
- **Federated Learning:** A technique where AI models are trained across decentralized data sources, so patient data remains on local servers [10] and is not shared with a central system.
- **Homomorphic Encryption:** A method [11] that enables data to be encrypted even while being processed, allowing AI to perform computations on encrypted data without compromising security [12].
- **Secure Multi-party Computation (MPC):** How multiple entities can jointly compute a function over their inputs without revealing any individual input [13].
- **De-identification and Anonymization**: Ensuring that personal identifiers are removed from datasets [14] while preserving the value of data for AI algorithms.

## 2.3 Regulatory and Ethical Considerations in AI-Driven Healthcare

AI in healthcare, also called Software as a Medical Device (SaMD), can significantly improve health outcomes but brings regulatory challenges. Regulations for AI are still developing, even in advanced regions. The European Commission's proposed AI Act aims to ensure safety while encouraging innovation. In the U.S., HIPAA safeguards patient information and highlights the importance of privacy [15].

In India, the National Health Policy (2017) [16] and the National Digital Health Blueprint (NDHB 2019) focus on digital health integration.It focuses on creating a system of electronic health records that adhere to international standards. It also establishes clear data ownership guidelines while incorporating ethical practices such as data anonymization and de-identification to ensure privacy and security [17]. The Digital Information Security in Healthcare Act (DISHA) 2018 seeks to protect electronic health data [18]. The General Data Protection Regulation (GDPR) sets strict data protection rules in Europe, and India's IT Act addresses electronic data security. Regulatory frameworks help evaluate the benefits and risks of AI products. The Medical Device Rules, 2017, define software as medical devices. However, there is a need for better standards for testing and validating AI technologies to ensure they are safe and effective.

## 3. Privacy-Preserving AI in Healthcare: Federated Learning and Differential Privacy for Secure Diabetes Prediction

In this case study, we implement a privacy-preserving approach to diabetes prediction using federated learning and differential privacy techniques. The dataset used is the Pima Indians Diabetes Dataset, which contains features related to diabetes diagnosis. We aim to demonstrate how decentralized machine learning models can be trained across multiple clients while ensuring that patient data remains secure.

### 3.1 Federated Learning Setup

We simulate a federated learning environment with **three clients**, where each client locally trains an XGBoost classifier on a subset of the dataset. The data is stratified and split to ensure that each client gets a balanced dataset representing the entire population. The clients use **SMOTE** (Synthetic Minority Over-sampling Technique) to handle the class imbalance in the diabetes dataset. Each client trains its model independently without sharing raw data, and only model updates are shared. After training, the individual models from each client are aggregated using a **weighted F1-score and data size** approach. This ensures that the global model performance reflects the contributions of each client while maintaining fairness and accuracy.

### 3.2 Data Encryption for Security

To protect model updates during communication between clients and the central server, we implemented **encryption** using the **Fernet** encryption scheme. After each local training round, the trained model parameters are encrypted before being shared with the central server. This prevents unauthorized access to sensitive model data that could otherwise reveal private information. Each client's model is encrypted using a randomly generated key, and only the encrypted model is shared. On the central server, the encrypted models are decrypted for further aggregation and analysis. This encryption layer adds an extra security measure, ensuring that even if the model is intercepted, it cannot be deciphered without the key.

### 3.3 Differential Privacy in Prediction

In addition to federated learning and encryption, we introduce **differential privacy** to further protect the privacy of individual patients during the prediction phase. **Noise** is added to the final model predictions to ensure that no single patient's data can be reconstructed or inferred from the predictions. By incorporating differential privacy, we make it harder for attackers to extract sensitive information, even from aggregated data. In our experiment, a small amount of noise is added to the predictions, ensuring a balance between privacy protection and model accuracy. The results show that the model maintains strong performance (with an accuracy of over 80%) while providing privacy guarantees for patient data.

### 3.4 Results and Evaluation

The final predictions are aggregated across clients using the weighted method described earlier. The results show that the federated model achieves a high accuracy of **84%** when evaluated on the test dataset. When applying differential privacy, the accuracy slightly decreases due to the added noise, but it remains within acceptable limits. Furthermore, the encrypted models show that privacy and security can be enhanced without sacrificing significant performance. This demonstrates the potential of combining **federated learning**, **encryption**, and **differential privacy** to create robust, privacy-preserving AI systems in healthcare.

Table 1. Experimental Results

| Model | Accuracy | F1 Score | Impact of DP | Impact of Encryption |
|---|---|---|---|---|
| Client 1 | 0.81 | 0.75 | Improved | Minimal |
| Client 2 | 0.88 | 0.84 | Improved | Minimal |
| Client 3 | 0.77 | 0.7 | Improved | Minimal |
| Global Model | 0.8312 | 0.82 | Improved | Minimal |
| Non-private Model | 0.71 | 0.62 | N/A | N/A |

Adding DP slightly reduces the global model's accuracy, likely due to the noise introduced by DP, but the reduction is minimal (~2%).All client models show a similar pattern of improvement over the rounds, with a slight drop in accuracy when DP is applied. The difference between models with and without DP varies from 1-3%, with Client 2 being the most robust to the effects of DP.

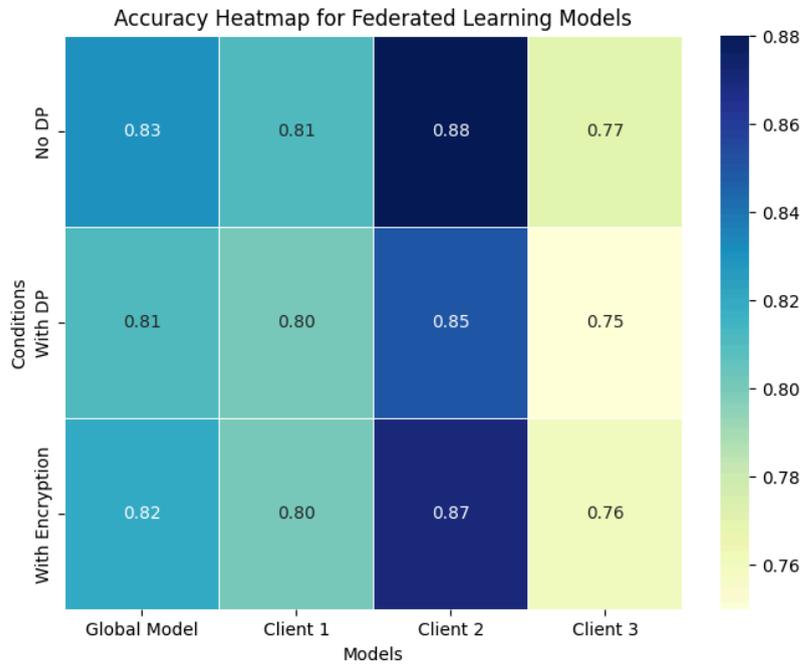

**Fig. 1.** Final Comparison of Model Accuracy in Federated Learning with differential privacy and encryption.

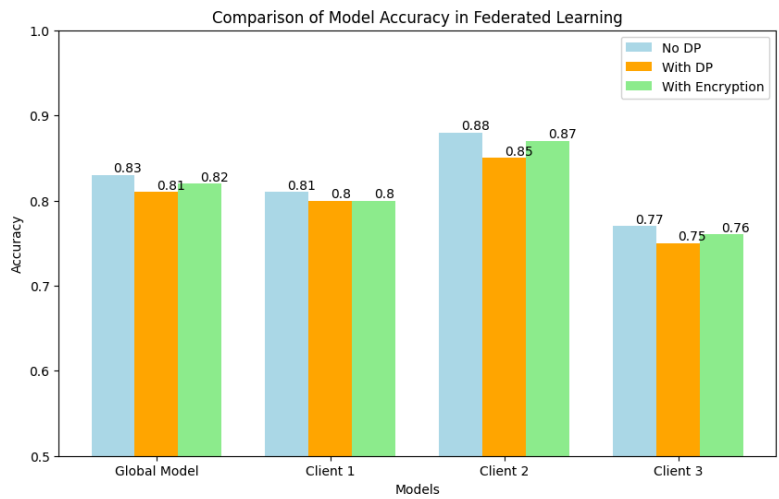

**Fig. 2.** Accuracy Heatmap for Federated Learning Models Across Conditions.

## 4.    Future Directions in AI, Security, and Privacy in Healthcare

The future of AI in healthcare will prioritize security and privacy. It will utilize federated learning to train AI models on decentralized data, keeping sensitive information secure. Advanced encryption methods, such as homomorphic encryption, will process encrypted data while protecting patient privacy. Explainable AI (XAI) will help healthcare professionals and patients understand AI decisions better. Additionally, blockchain technology may enhance data security and integrity. These developments aim to improve healthcare outcomes while safeguarding sensitive patient data. Ongoing research and collaboration among different groups will be important for creating strong AI solutions that follow ethical guidelines. It will also be essential to train healthcare staff on how to use AI tools effectively. These efforts aim to enhance healthcare outcomes while protecting sensitive patient information. By tackling these issues, AI can make healthcare more efficient, secure, and focused on patient needs.

## 5.    Conclusion

In summary, the paper has explored the critical balance between security and privacy in the application of AI within healthcare systems. The implementation of AI models must consider ethical and regulatory frameworks that protect patient data while still harnessing the power of advanced analytics. The code implementation provided demonstrates practical methods for maintaining privacy and security through techniques such as federated learning and differential privacy. As healthcare systems grow, it's important for stakeholders to focus on secure and privacy-protecting AI technologies. Doing so will build patient trust and improve healthcare delivery. This will help ensure that AI is used responsibly and effectively for better health outcomes in the future.